\pdfoutput=1
\documentclass[12pt]{article}

\usepackage{sbc-template}
\usepackage{graphicx,url}
\usepackage[latin1]{inputenc}
\usepackage{subfigure}
     
\title{Representation of Evolutionary Algorithms in FPGA Cluster for Project of Large-Scale
	Networks}

\author{Andre B. Perina\inst{1}, Marcilyanne M. Gois\inst{2}, Paulo Matias\inst{3}, Joao M. P.
	Cardoso\inst{4}, \\ Alexandre C. B. Delbem\inst{1}, Vanderlei Bonato\inst{1}}

\address{Institute of Mathematical and Computer Sciences \\
	University of São Paulo (USP) -- São Carlos, SP -- Brazil
	\nextinstitute
	São Carlos School of Engineering \\
	University of São Paulo (USP) -- São Carlos, SP -- Brazil
	\nextinstitute
	São Carlos Institute of Physics \\
	University of São Paulo (USP) -- São Carlos, SP -- Brazil
	\nextinstitute
	Faculty of Engineering \\
	University of Porto (UP) -- Porto -- Portugal
	\email{\{abperina,mmgois,paulo.matias\}@usp.br,jmpc@acm.org,\{acbd,vbonato\}@usp.br}}

\begin{document} 

	\maketitle

	\begin{abstract}
		Many problems are related to network projects, such as electric distribution,
		telecommunication and others. Most of them can be represented by graphs, which
		manipulate thousands or millions of nodes, becoming almost an impossible task to obtain
		real-time solutions. Many efficient solutions use Evolutionary Algorithms (EA),
		where researches show that performance of EAs can be substantially raised by using an
		appropriate representation, such as the Node-Depth Encoding (NDE). The objective of this
		work was to partition an implementation on single-FPGA (Field-Programmable Gate Array)
		based on NDE from 512 nodes to a multi-FPGAs approach, expanding the system to 4096 nodes.
	\end{abstract}

	\section{Introduction}

		Nowadays, it is not so difficult to find problems that can be mapped to a network in order
		to improve solutions for power distribution networks, highways, etc. All these problems can
		be modeled as graphs, and many times the solution of the problem is based on finding the
		Minimum Spanning Tree (MST) of the modeled graph. It must be noted that normally
		these graphs are extense and complex, leading to complex computational systems with high
		time cost. Many efficient solutions for these problems are based on Evolutionary Algorithms
		(EA), which showed to have good perfomance and efficiency \cite{DE2006}.

		Finding a MST is solvable in polynomial time by using two well-known
		algorithms: Prim and Kruskal. However, these algorithms fail
		when a degree restriction is applied to all nodes of a graph, which is not an uncommon
		cenario on real-world problems. In this case, the problem becomes NP-Hard \cite{KNOWLES2000}.

		In this paper we propose an expansion to a platform developed by \cite{GOIS2014} in order to solve bigger problems, by
		exploiting the use of multi-FPGAs. We present a framework for modeling and then a real
		implementation on hardware.

	\section{Materials and Methods}

			In a previous work, \cite{GOIS2014} developed a platform called NDEWG, which finds a MST
			for a given graph by using an EA with complexity of $\mathrm{O(}\sqrt{n}\mathrm{)}$
			\cite{DELBEM2012} with serial processing elements, or near
			$\mathrm{O(}\frac{\sqrt{n}}{\sqrt{n}}\mathrm{)}=\mathrm{O(}1\mathrm{)}$ if $\sqrt{n}$
			processing elements (henceforth called Workers) work in parallel. This complexity
			was achieved by using a special representation for trees, called Node-Depth Encoding, or NDE \cite{SANTOS2010}.

			The NDE basically represents a tree on a linear list
			containing a pair of values ($n_x$, $d_x$), where $n_x$ are the tree nodes and $d_x$
			their depths. The pairs are disposed in the list based on the depth search algorithm
			traversing.

			The NDEWG hardware transforms a graph into
			a forest of spanning trees by using a modified version of the Kruskal algorithm. A pair of trees is randomly chosen and a tree operator called Preserve
			Ancestor Operator (PAO) \cite{SANTOS2010},\cite{LIMA2008} is applied to both, generating
			two different but still spanning trees from the original graph, respecting any applied
			restriction degree. The results are analised for any improvements in overall weight, and
			discarded if no optimisation is achieved.
			Many Workers may apply in parallel the PAO operator with different random seeds and each
			analyses for best weight reduction. After many iterations, the result is an
			optimised spanning tree with degree restriction.

			This project was implemented on an Altera Cyclone II FPGA, running up to 512 nodes and
			later on a Stratix IV GX, reaching up to 1024 nodes. Higher expansions weren't possible
			due to bus-width restrictions. To reach a higher size of solvable graphs, the system bus
			was expanded to 64 bits.

\begin{figure}[ht]
				\centering
				\includegraphics[width=.6\textwidth]{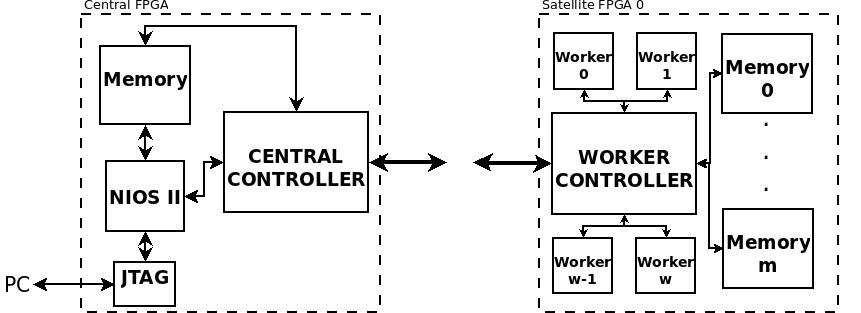}
				\caption[]{Central FPGA (left) and Sat. FPGA (right) organisations}
				\label{fig:edndewg}
			\end{figure}

			Since most of the hardware resources of the platform are dedicated to Worker Synthesis,
			which increases linearly with the graph size, the Workers were moved to separate set of
			FPGAs, enabling scalability. The proposed platform was a star architecture consisted by one Central FPGA, responsible
			for managing the spanning trees to be worked by the Workers and many Satellite FPGAs,
			each having a number of Workers. Each FPGA communicates with the Central FPGA by using a
			dedicated link. Figure~\ref{fig:edndewg} shows how each FPGA logic was implemented.

			For the validation on physical hardware, it was still necessary to develop a network system
			for connecting the FPGAs. We connected the systems by using high-speed 10 Gbps Ethernet Interfaces. A whole
		system was developed to abstract all the network layer aspects. This system was called Network Abstraction System, or NAS.

		The NAS implements master and slave Avalon Memory-Mapped frontends, which are compatible
		with the network ports of the previously Central and Satellite modules. The system converts
		the transactions to a streaming-based Avalon bus, which is used to communicate with the
		Ethernet MAC. The Ethernet MAC and XAUI PHY operates on the physical network layer.

\begin{figure}[ht]
			\centering
			\includegraphics[width=.85\textwidth]{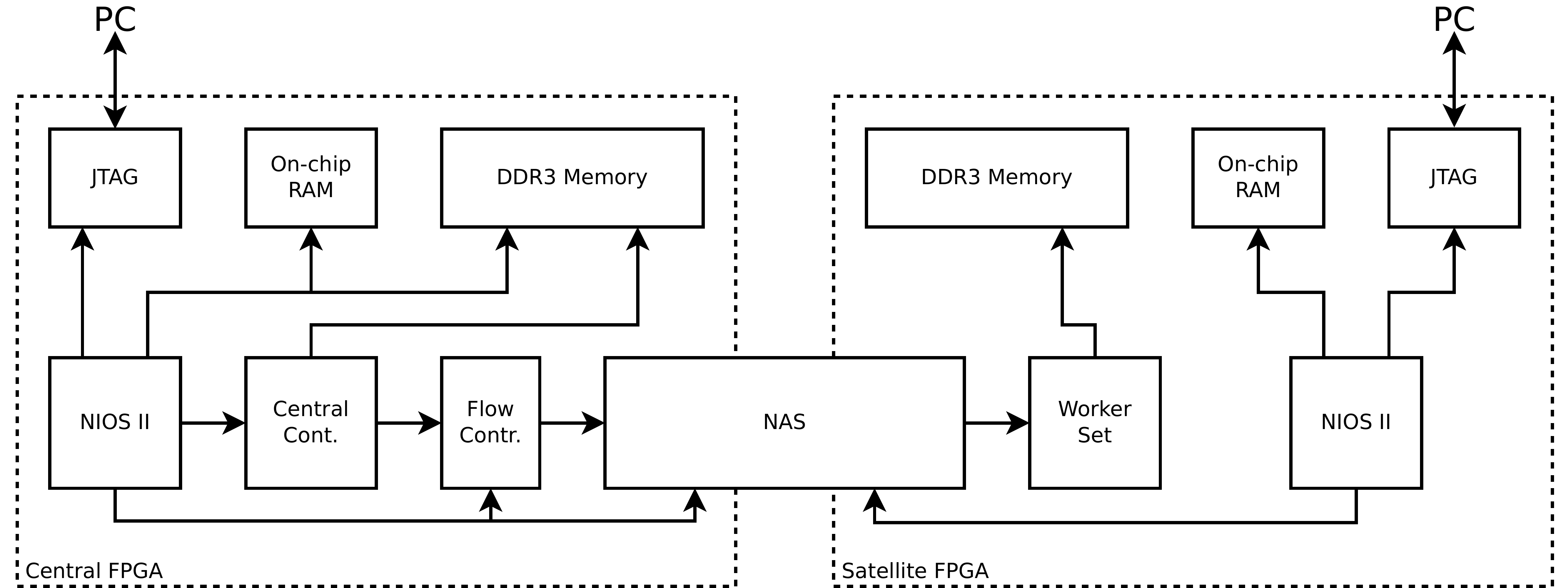}
			\caption[]{Final Validation Architecture}
			\label{fig:finalsystem}
		\end{figure}

		The Central, Satellite and NAS projects were merged to validate the whole system.
		It was used two Stratix V GX Development Kits (5SGXEA7) connected using two DUAL XAUI TO
		SFP+ HSMC BOARD, and two full-duplex SFP+ Avago AFBR-703SDZ optical interfaces, with speed
		up to 10 Gbps. Figure~\ref{fig:finalsystem} shows the final validation system.

	\section{Results}

		The Central and Satellites modules were first tested by simulation, where no memories and
		interconnection delays were considered. Table~\ref{tab:model} shows some results from
		simulation and synthesis tools, where one local memory was used for each Sat. module, $c^c_\mathrm{elem}$ and 
		$c^s_\mathrm{elem}$is the amount of Logic Elements (ALUTs) used on the Central and
		Satellites modules respectively, $f^c_\mathrm{max}$ and $f^s_\mathrm{max}$
		the maximum frequency (MHz) for Central and Sat. modules, and the average time per
		iteration (in seconds) to solve graphs up to 1024 nodes.

		\begin{table}[ht]
			\centering
			\caption{Results for Central/Sat Simulation Model}
			\label{tab:model}
			\begin{tabular}{|c|c|c|c|c|c|c|c|}
				\hline
				Mem. & Sat. FPGAs & $c^c_\mathrm{elem}$ & $c^s_\mathrm{elem}$ &
					$f^c_\mathrm{max}$ & $f^s_\mathrm{max}$ & Avg. Time (per Iter.) & Speedup \\
				\hline
				1 & 1 & 1552 & 56352 & 174.4 & 75.91 & 1.44e-07 & 1x \\
				1 & 4 & 2945 & 13956 & 124.42 & 122.43 & 8.34e-08 & 1.73x \\
				1 & 8 & 5051 & 7142 & 122.99 & 123.92 & 9.43e-08 & 1.53x \\
				\hline
			\end{tabular}
		\end{table}

\begin{figure}[ht]
			\centering
			\includegraphics[width=.5\textwidth]{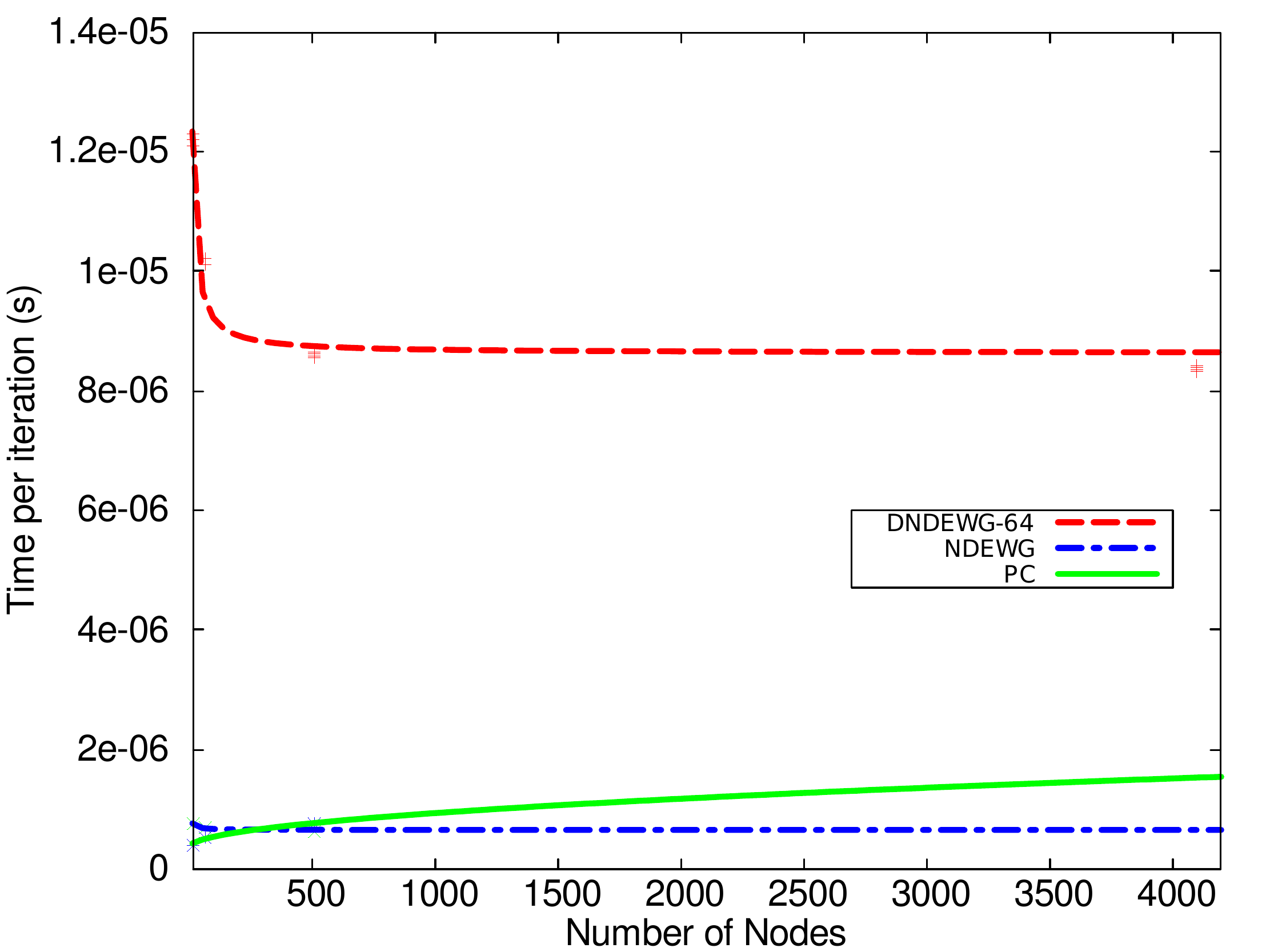}
			\caption[]{Average processing time between systems}
			\label{fig:exectime}
		\end{figure}

		On the final hardware platform, the size of solvable graphs was upgraded to 4096 nodes,
		since expansion to 8192 nodes was not possible due to lack of available on-chip memory.
		Figure~\ref{fig:exectime} shows the average processing time on DNDEWG-64 (multi-FPGAs approach), NDEWG-32 (previous work) and on a
		Intel Core 2 Quad Q6600 (2.40 GHz) with the operator coded in C, running in parallel by using
		OpenMP. Data for NDEWG-32 and PC are available only up to 512 nodes.

		The NDEWG shows a better efficiency, since it is not affected by a latency generated by the
		network interface. However its scalability is strongly held by its monolithic nature, restricted
		to resources of only one FPGA.
		For 1024 nodes, the DNDEWG-64 has an average of approx. 85 us, whereas in simulation (1 memory,
		1 Sat. module), the average time was of 0.144 us. Thus the bottleneck is concentrated on the
		network system and must be investigated for optimisations.

	\section{Conclusion}

		Parallelised
		Evolutionary Algorithms are good candidates for optimising NP-Hard problems. This kind of
		solution can be applied to real world problems, such as electric distribution,
		where complex networks can have up to 100,000 nodes.
		Much work must still be done in order to improve the multi-FPGAs approach. The NAS works on a
		low speed compared to the high-speed link, hence improvements on the system must be made and
		are objectives for further research.

\bibliographystyle{sbc}
\bibliography{sbc-template}

\end{document}